\documentclass[aip,reprint]{revtex4-1}

\usepackage{graphicx}
\usepackage{amsmath}
\usepackage{xcolor}
\usepackage[utf8]{inputenc}
\usepackage[T1]{fontenc}
\usepackage[colorinlistoftodos]{todonotes}
\usepackage{multirow}
\usepackage{CJKutf8}
\usepackage{lineno}

\newcommand{\gd}{\ensuremath{^{148}\mathrm{Gd}}}

\newcommand{\YAP}{YAP:Ce}
\newcommand{\VF}{\ensuremath{V_\mathrm{F}}}
\newcommand{\dd}{\ensuremath{\mathrm{d}}}
\newcommand{\kB}{\ensuremath{kB}}
\newcommand{\Gd}{\ensuremath{^{148}\mathrm{Gd}}}
\newcommand{\Cs}{\ensuremath{^{137}\mathrm{Cs}}}
\newcommand{\Co}{\ensuremath{^{60}\mathrm{Co}}}
\newcommand{\B}{\ensuremath{^{10}\mathrm{B}}}
\newcommand{\Li}{\ensuremath{^{7}\mathrm{Li}}}
\newcommand{\LANL}{Los Alamos National Laboratory, Los Alamos, New Mexico, 87545, USA}
\newcommand{\UKy}{University of Kentucky, Lexington, Kentucky, 40506, USA}
\newcommand{\NCSU}{North Carolina State University, Raleigh, North Carolina 27695, USA}
\newcommand{\Caltech}{W. K. Kellogg Radiation Laboratory, California Institute of Technology, Pasadena, California 91125, USA}
\newcommand{\TTU}{Tennessee Technological University, Cookeville, Tennessee 38505, USA}

\newcommand{\UIUC}{University of Illinois, Champaign, Illinois 61820, USA}
\newcommand{\ETSU}{East Tennessee State University, Johnson City, Tennessee 37614, USA}

\begin{document}


\title{YAP:Ce scintillator as an absolute ultracold neutron detector} 



\author{M. Krivo\v{s}}
\email{mkrivos@lanl.gov}
\author{Z. Tang (汤兆文)}
\affiliation{\LANL}
\author{N. Floyd}
\affiliation{\LANL}
\affiliation{\UKy}
\author{C. L. Morris}
\affiliation{\LANL}
\author{M. Blatnik}
\affiliation{\LANL}
\affiliation{\Caltech}
\author{C. Cude-Woods}
\affiliation{\LANL}
\affiliation{\NCSU}
\author{S. M. Clayton}
\affiliation{\LANL}
\author{A. T. Holley}
\affiliation{\TTU}
\author{T. M. Ito}
\affiliation{\LANL}
\author{C.-Y. Liu}
\affiliation{\UIUC}
\author{M. Makela}
\author{I. F. Martinez}
\author{A. S. C. Navazo}
\affiliation{\LANL}
\author{C. M. O'Shaughnessy}
\affiliation{\LANL}
\author{E. L. Renner}
\affiliation{\LANL}
\author{R. W. Pattie}
\affiliation{\ETSU}
\author{A. R. Young}
\affiliation{\NCSU}


\date{\today}

\begin{abstract}
The upcoming UCNProBe experiment at Los Alamos National Laboratory will measure the $\beta$-decay rate of free neutrons with different systematic uncertainties than previous beam-based neutron lifetime experiments. We have developed a new \B-coated \YAP\ scintillator whose properties are presented. The advantage of the \YAP\ scintillator is its high Fermi potential, which reduces the probability for upscattering of ultracold neutrons, and its short decay time, which is important at high counting rates.  Birks' coefficient of \YAP\ was measured to be $(5.56 ^{+0.05}_{-0.30})\times 10^{-4}$~cm/MeV and light losses due to 120~nm of \B-coating to be about 60\%. The loss of light from \YAP\ due to transmission through deuterated polystyrene scintillator was about 50\%. The efficiency for counting neutrons that are captured on the \B\ coating is $(86.82~\pm~2.61)\%$. Measurement with ultracold neutrons showed that \YAP\ crystal counted 8\% to 28\% more UCNs compared to ZnS screen. This may be due to an uneven coating of \B\ on the rough surface.
\end{abstract}

\begin{CJK*}{UTF8}{gbsn}
\maketitle 
\end{CJK*}
\section{Introduction}
The idea of ultracold neutrons (UCN) dates back to Y. B. Zeldovich\cite{Zeldovich1959}, who estimated that with sufficiently low kinetic energy, neutrons can reflect off materials with positive Fermi potentials\cite{fermi1936} at all angles of incidence. The Fermi potential, \VF, sets a limit on the maximum neutron kinetic energy, below which losses due to neutron inelastic upscattering and absorption on the contact with a material are significantly reduced and the losses are on the order of $10^{-4}$ per bounce. This enables UCN to be trapped in storage containers for long periods of time and allows many measurements of the neutron properties. At this kinetic energy, the neutrons can also be manipulated easily by gravitational or magnetic potential energy. These properties allow many high precision measurements using the neutron, and many UCN sources\cite{Steyerl1975, Altarev1980,Steyerl1986,Saunders2013, Ito2018,Zimmer2011,Martin2021} have been built to exploit these properties.
UCN have been used, for example, in the measurements of the neutron decay lifetime\cite{Serebrov2005, Pichlmaier2010,Pattie2018, Ezhov2018, Serebrov2018,Gonzalez2021}, asymmetries in the neutron beta decay\cite{Pattie2009,Broussard2013}, neutron electric dipole moment\cite{GOLUB1994,Ito2018,Baker2006,Abel2020}, gravitational quantum states of the neutron.\cite{Nesvizhevsky2002,Pignol2007,Jenke2011}

The neutron lifetime, $\tau_n$, is measured in two ways: a beam and a bottle method.\cite{RevModPhys.83.1173} The former method uses a cold neutron beam and measures the number of decayed protons from Equation~(\ref{eq:neutron_decay}). \cite{Byrne1996, Nico2005,Yue2013,Nagakura2016,Hoogerheide2019} 
\begin{equation}
    n \rightarrow p + e + \bar{\nu}_{e}
    \label{eq:neutron_decay}
\end{equation}
The latter method traps UCN inside a volume and counts their numbers after different time intervals.\cite{Serebrov2005, Pichlmaier2010,Pattie2018, Ezhov2018, Serebrov2018,Gonzalez2021} There is an on-going discrepancy with a significance greater than $4.5\sigma$ between the two methods. A newly proposed experiment, UCNProBe, will measure the neutron beta decay lifetime by counting one of the decay production, the electron, to the 1-2 seconds level, which may help resolve this discrepancy. As with all beam-type experiments, UCNProBe  requires measurements of two absolute quantities, in this case, $\tau_n$ can be extracted by:
\begin{equation}
    \frac{1}{\tau_n} = \frac{\epsilon_\beta\dot{\beta}(t)}{\epsilon_N N(t)},
    \label{eq:lifetime}
\end{equation}
where $\dot{\beta}(t)$ is the electron detection rate, $N(t)$ is the numbers of neutrons available for decay at a given time $t$, and $\epsilon_i$ refers to the detection efficiencies for the decay product and the neutron. These efficiencies need to be known absolutely for the extraction of the lifetime to the 0.1\% level.

In the UCNProBe experiment, UCN will be filled into a 12.7~cm deuterated-Polystyrene (dPS) scintillator box. The dPS box will serve as a trap for UCN due to its high value of \VF \cite{Tang2021} and as in-situ counter for the decay electrons. To measure $N(t)$, a retractable dagger with a \B-coated scintillator will extend into the chamber and capture all the neutrons. One of the common ways to detect UCN is to utilize the large neutron capture cross section for $^{10}$B shown by the following process:

\begin{equation}
\begin{split}
n + {}\B \rightarrow & \, \Li + \alpha  \qquad\,\,  Q = 2792~\mathrm{keV} \quad(6\%)\\
n + {}\B \rightarrow & \, \Li^* + \alpha \qquad  Q = 2310~\mathrm{keV} \,\,\,(94\%)\\
                       & \, \Li^* \rightarrow {}\Li + 478~\mathrm{keV}
\end{split}
\label{eq:boron_capture}
\end{equation}
About 6\% of the reactions lead to a ground state of \Li\ and the remaining 94\% lead to an excited $\Li^*$ state followed by a 478~keV gamma. Energetic ions from this reaction are then detected in a scintillator. Due to the large neutron capture cross section and the low velocity of the UCN, only 120~nm of \B\ is needed to achieve a capture probability 96\% for 4 m/s neutrons. The standard scintillator used for this method of UCN counting measurements is a ZnS:Ag screen. \cite{Wang2015} Since ZnS is a powdered scintillator pressed onto an adhesive layer, it is difficult to characterize the UCN upscattering probability; this prevents direct correlation between the UCN capture counts on \B\ with the absolute numbers of UCN in the dPS box to the desired precision level of 0.1\%. 
Therefore, the search for a new scintillator that has a high Fermi potential and good light output for heavy ions is needed. 

This paper presents a  study of a Cerium-activated Yttrium Alluminum Perovskite (\YAP) scintillator. \YAP\ has a calculated \VF\ of 148~neV\ and a scintillation decay time of 28~ns.\cite{YAP_Epic_crystal,crytur_yap} In contrast with ZnS, this scintillator can be fabricated in crystal form, which means that neutrons that aren't captured will simply bounce back into the volume and will be counted on their next interaction. Thus, the systematic uncertainties associated with upscattering can be controlled at $\sim 10^{-4}$ level. This allows us to determine the neutron detection efficiency by tagging the 478~keV gamma with the charged particle detection in the \YAP\ scintillator. The short decay time of \YAP\ will also reduce the counting dead time. For comparison, ZnS has a complicated spectrum with tails that go out to many $\mu$s.

\section{Scintillation properties of YAP}
\begin{figure}
    \centering
    \includegraphics[width=\linewidth]{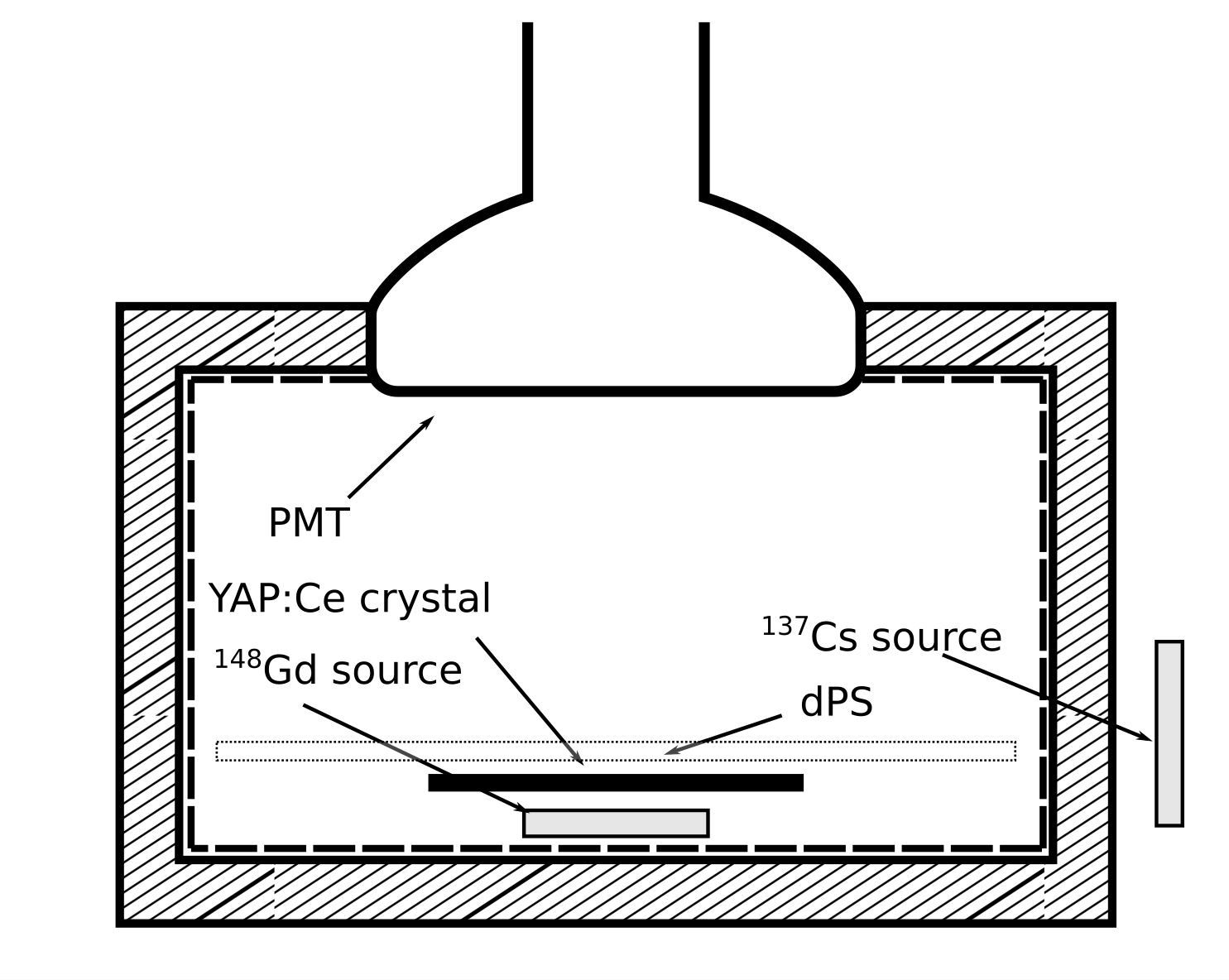}
    \caption{A side-view cross-section of the stainless steel chamber. The black rectangle represents the \YAP\ crystal and grey rectangles are \gd\ and \Cs\ sources. The dotted rectangle above \YAP\ shows the placement of the dPS scintillators. The Teflon lining is shown as a dashed line inside the chamber.}
    \label{fig:exp1_cartoon}
\end{figure}

Scintillation properties of \YAP\ were evaluated without the UCN, and are summarized below. These experiments were performed in a stainless-steel chamber as is shown in Figure~\ref{fig:exp1_cartoon}. The chamber was lined with about 2~mm thick Teflon sheets\cite{Teflon_TFCO}, so it could serve as $4\pi$-integrating-sphere. A 78 mm ET-9305QB photomultiplier tube (PMT) with fused silica window was mounted on top of the chamber. The output from the PMT was processed directly using a CAEN V1730 waveform ADC with 2~ns sampling rate.  For the results, 992~ns long digitized waveforms were saved and analyzed. For each waveform, a constant offset calculated from 100~ns before the trigger was subtracted. A pre-trigger of 120~ns ensured that this offset does not contain the signal from \YAP. Three measurements were completed in the chamber. Firstly, the light output for $\gamma$ and $\alpha$ particles incident on the scintillator were measured, and the Birks' coefficient\cite{Birks:1951} was extracted. Secondly, due to possibility of light attenuation from the \B-coating, a comparison between a plain \YAP\ scintillator and a \B-coated version was performed. Thirdly, the absorption and re-emission of \YAP's light by dPS was measured. In the final UCNProBe experiment, scintillation light from \YAP\ must pass through the dPS scintillator box before being collected by PMTs, which may cause losses on \YAP's light output on the dPS. All three measurements are presented in this section.

\begin{figure}
    \centering
    \includegraphics[width=\linewidth]{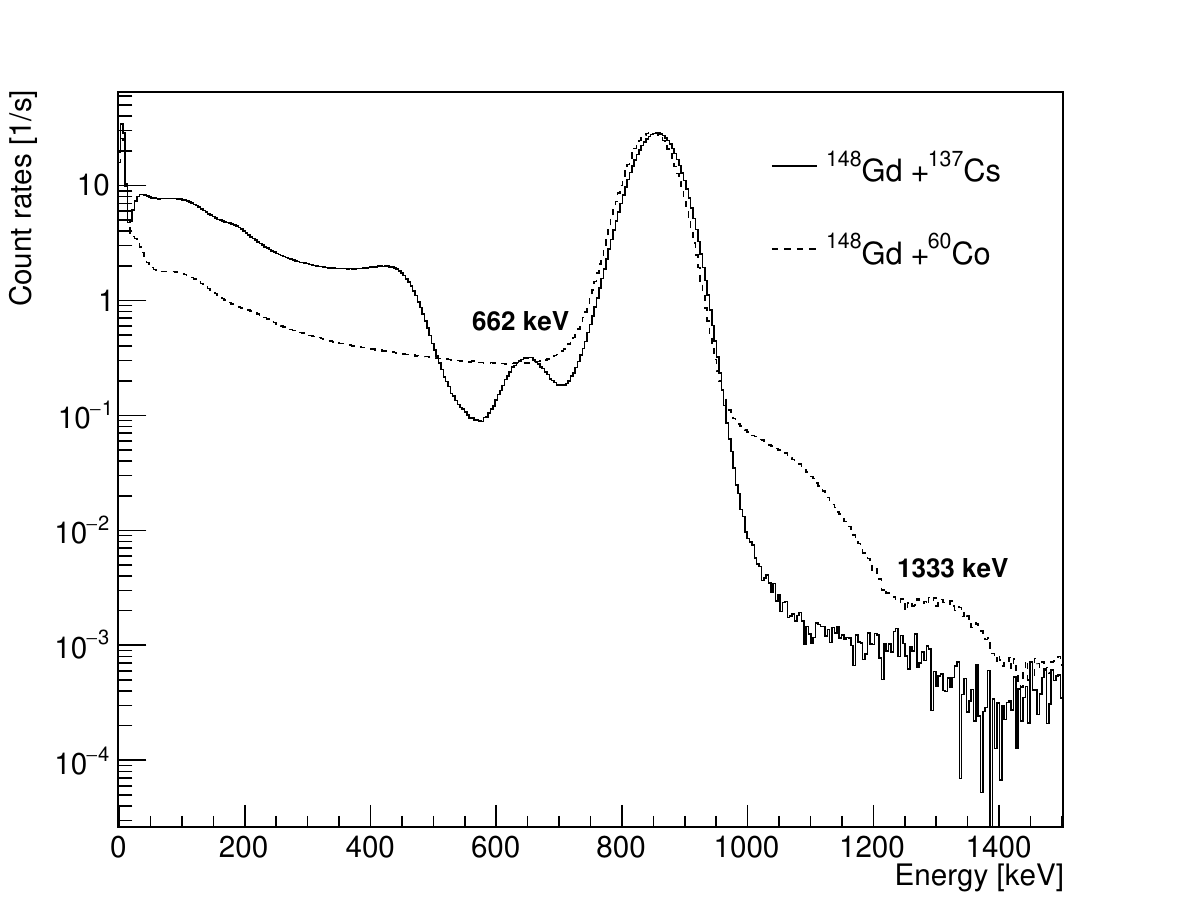}
    \caption{Count rates for a measurement of \Gd\ and \Cs\ (solid line) and \Gd\ and \Co\ (dashed line).}
    \label{fig:GdCsCo}
\end{figure}

\begin{figure}
    \centering
    \includegraphics[width=\linewidth]{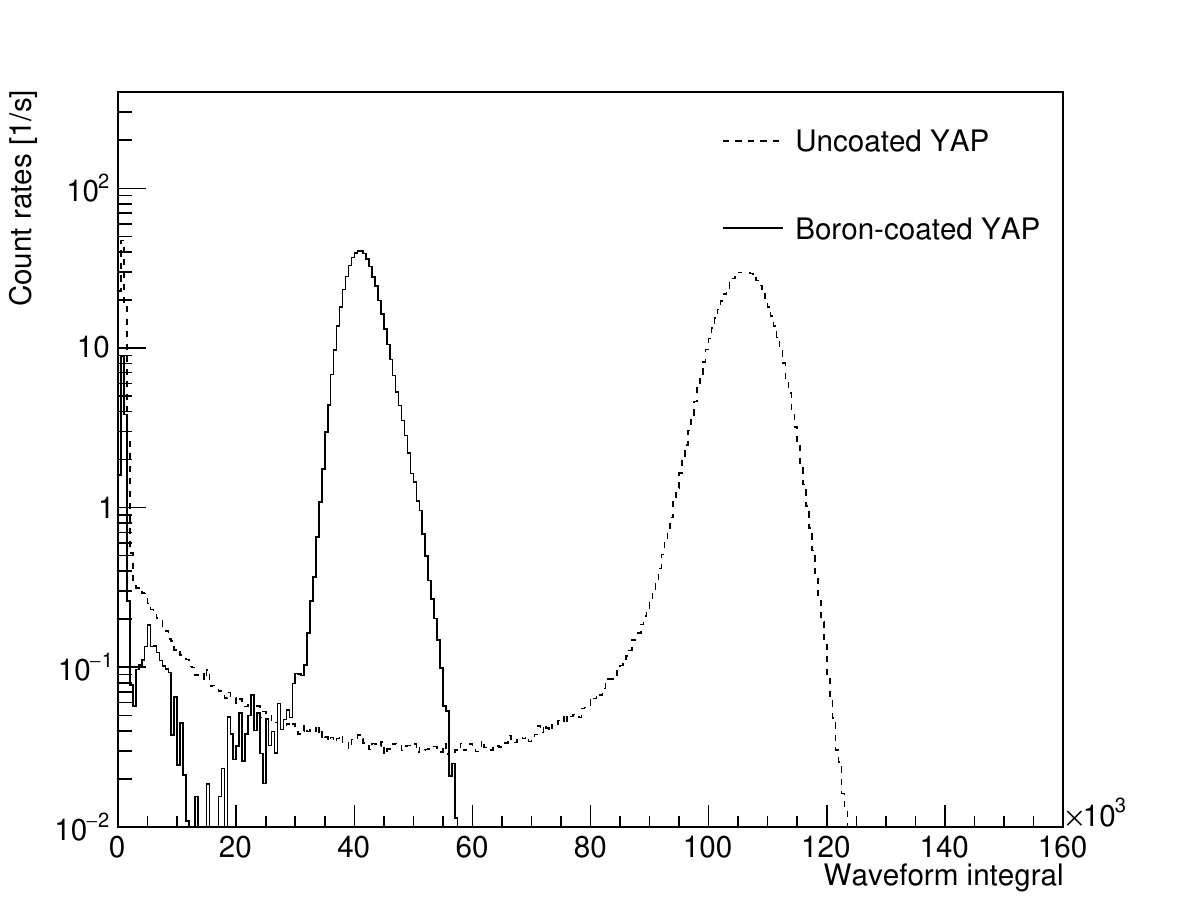}
    \caption{A comparison of count rates for \Gd\ source for uncoated \YAP\ scintillator (dashed line) and the \YAP\ scintillator with a 120 \B\ coating from both sides (solid line).}
    \label{fig:coated_YAP}
\end{figure}

The light output for heavy charged particles from the \YAP\ scintillator was measured and assumed to follow Birks' law\cite{Birks:1951,Knoll:2010xta}:
\begin{equation}
    L(\kB) \propto  \int^{E_\alpha}_{0}\frac{\frac{\dd E}{\dd x}}{1 + kB\frac{\dd E}{\dd x}}\dd x,
    \label{eq:birks_law}
\end{equation}
where $E_\alpha$ is the energy of the $\alpha$ particle, $L(\kB)$ is the light output from the scintillator, $\dd E/\dd x$ is the characteristic energy loss, and \kB\ is Birks' coefficient. To estimate \kB, light output from a known $^{148}$Gd source was used.\cite{NuDat2024} For energy calibration, gamma peaks of 662~keV from $^{137}$Cs and 1333~keV of $^{60}$Co were used. 
To minimize PMT gain shifts, two measurements were performed with the PMT turned on for the whole duration. The $^{148}$Gd electroplated source 2.54~cm in diameter was inserted into the chamber as shown in Figure~\ref{fig:exp1_cartoon}, and the \YAP\ scintillator was placed directly above it. To minimize the $\alpha$-particle losses in air, the chamber was evacuated to a pressure of below $10^{-4}$~Torr.
First, a \Cs\ source was placed next to the chamber and both \Gd\ and \Cs\ spectra were taken. With the PMT still on and the chamber untouched, the \Cs\ source was replaced with the \Co\ source and a second spectrum was measured. Both spectra with indicated full-energy peaks are shown in Figure~\ref{fig:GdCsCo}. A spectrum consisting of only \Gd\ source was measured as well, and is shown as uncoated \YAP\ in Figure~\ref{fig:coated_YAP}. To subtract the background, a separate run without the sources was also taken after power cycling the PMT power supply.
In the region-of-interest, i.e. near the \Gd\ peak, the background contribution was less than 0.01\%.

An energy calibration was applied by using the full-energy peaks of \Cs\ and \Co\ sources. The systematic uncertainty due to gain shifts was determined by calculating the \kB\ coefficient for \Gd\ peak positions for \Cs\ and \Co\ source separately and taking its weighted average. The \Gd\ $\alpha$-source has a tabulated energy of $3182.69$~keV. Two effects can, however, decrease this energy: one is a non-scintillating \textit{dead-layer} present on the surface of the scintillator and the other is an adhered layer of material on the source itself. A recent study\cite{Viererbl2016} showed that \YAP\ has a very thin dead-layer below 200~nm. If we model the dead-layer as non-scintillating thin layer where the alpha-particle loses energy, a 200~nm dead layer in \YAP\ results in an effective $73$~keV drop in energy for normal incidence. The second effect was measured to result in about $25$~keV decrease in energy for $\alpha$-particles.\cite{Nab_communications} The central value of the result was calculated with the tabulated \Gd\ $\alpha$-particle energy and a systematic uncertainty was added by repeating the analysis procedure, but with a lower $\alpha$-particle energy; this contributed only to one side of the final uncertainty.

The calculated $\alpha$-particle light-output of the \Gd\ source was $(886.18 \pm 6.16)$~keV, which is 27\% of the tabulated value. The corresponding Birks' coefficient was evaluated to be $(5.56 ^{+0.05}_{-0.30})\times 10^{-4}$~cm/MeV. The detail of uncertainties is shown in Table~\ref{tab:kB_uncert}. The best $kB$ result was compared to the $^{226}$Ra decay spectrum shown in Figure~5 in Ref.~\onlinecite{Baryshevsky1991}. Assuming a linear relation between the light output and a channel number $L=kN_\mathrm{ch}$, the calibration constant, $k$, was consistent within 3\% for individual peaks. For comparison, substituting $L$ for $\alpha$-particle energy $E_\alpha$ and assuming linear energy losses (i.e., $kB=0$), the calibration constants were consistent at the 10\% level.

\begin{table*}[htbp]
\begin{center}
\begin{tabular}{ccccccc}
Quantity & Value: & Dead layer & Integration time & Statistical & \multicolumn{2}{c}{Total}  \\ 
\hline
$kB$ [$10^{-4}$cm/Mev] &5.56 & $0.29$ & $0.045$ & $0.03$ & $+0.0541$ & $-0.3$ \\ 
$L$ [MeV] & 886.181 & N/A & 5.17 & 3.35 & \multicolumn{2}{c}{$\pm6.16$} \\
\end{tabular}
\end{center}
\caption{Detail of uncertainties for Birks' coefficient $kB$ and the light output from \Gd\ $\alpha$-particle. The uncertainty due to the dead layer affects only the lower bound of $kB$ and is not applicable to $L$.}
\label{tab:kB_uncert}
\end{table*}

To quantize the light losses on the \B\ layer, a \Gd\ $\alpha$-particle spectrum was measured using the uncoated version of \YAP\ (used in the above measurement) and a second sample of \YAP, which had 120~nm of \B-coating on both sides. Resulting count rates are shown in Figure~\ref{fig:coated_YAP}. Assuming a linear energy calibration, light losses due to the \B\ layer can be estimated by a shift between the two peaks. By fitting both peaks to a Gaussian function and comparing their means, a light loss of about 60\% was determined.

\begin{figure}
    \centering
    \includegraphics[width=\linewidth]{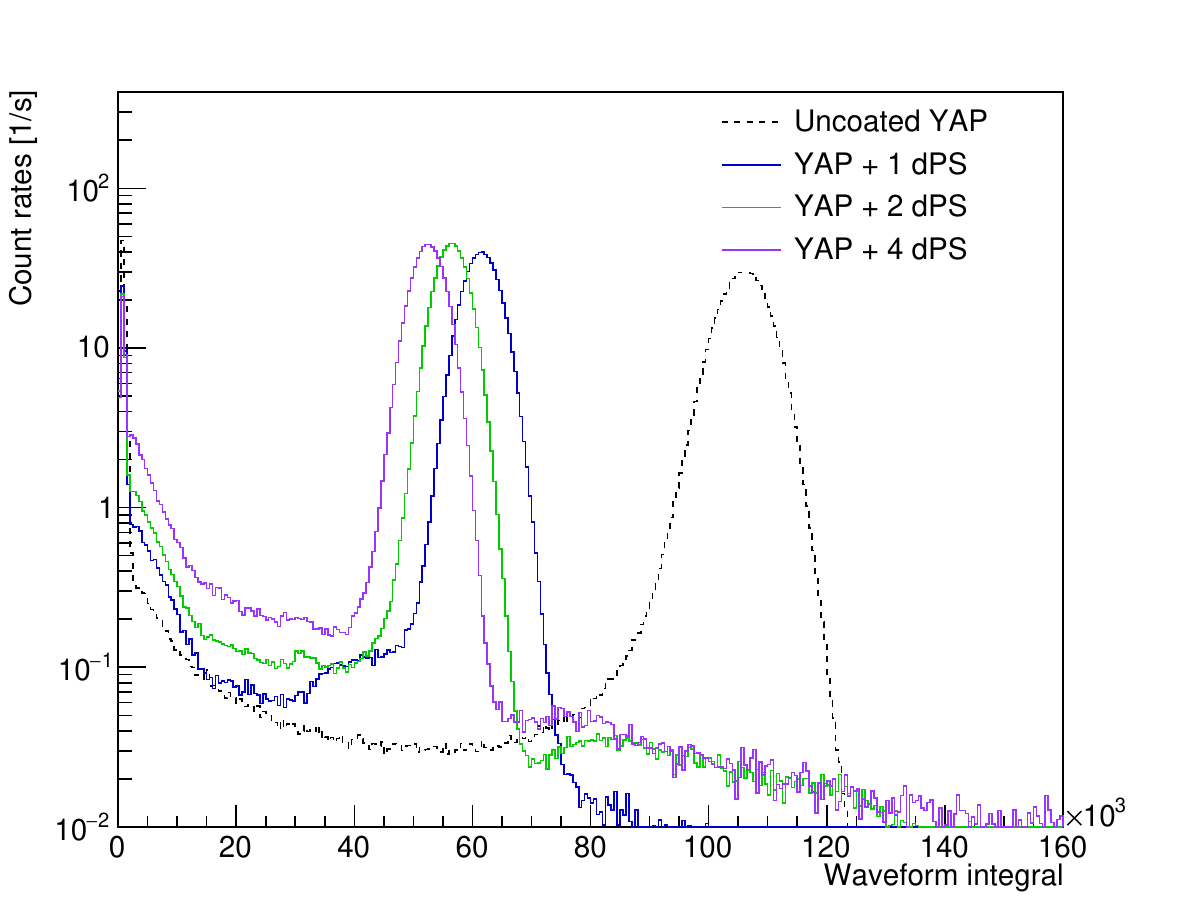}
    \caption{A comparison of count rates for \Gd\ source between uncoated \YAP\ scintillator by its own (dashed line) and for uncoated \YAP\ with one, two and four layers of dPS scintillator (solid blue, green, and violet, respectively).}
    \label{fig:YAP_dPS}
\end{figure}

Since the light emission spectrum of YAP overlaps with the absorption spectrum of the plastic scintillator used in the UCNProBe experiment (Eljen-299-02)\cite{Tang2021,floyd2023scintillation}, a separate set of  measurements were conducted to study the light attenuation due to these plastic scintillators. Figure~\ref{fig:exp1_cartoon} shows the arrangement in the chamber. Measurements with uncoated \YAP\ itself and with the addition of one, two, and four pieces of PS scintillators are shown in Figure~\ref{fig:YAP_dPS}. Again, by assuming linear energy calibration, a light loss of about 50\% was measured for one PS scintillator with much less further decrease with more sheets. 

\section{Experiments with UCN}
To establish \YAP\ as a viable absolute UCN counting scintillator, measurements with UCN were conducted at the Los Alamos UCN source \cite{Saunders2013, Ito2018}. A coincidence measurement with 478~keV photons from \B\ capture (Equation~(\ref{eq:boron_capture})) and a comparison with ZnS screen are presented in this section. 

\begin{figure}
    \centering
    \includegraphics[width=\linewidth]{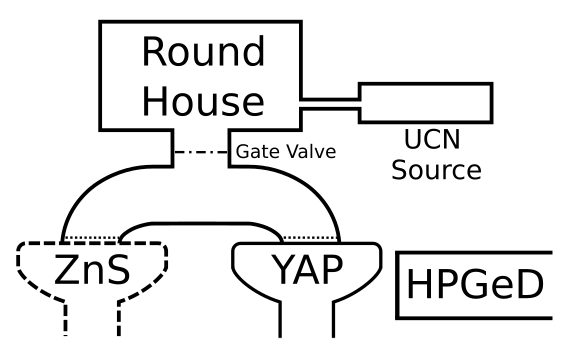}
    \caption{A schematic of two 51~mm PMTs attached to an UCN port with a high-purity germanium detector is shown as well. The second PMT with ZnS scintillator is shown with a dashed line, as it will be removed and replaced for a stainless-steel end-cap when measuring the 478~keV gamma coincidences. Round house is a large volume container used to smooth-out a UCN delivery from the spallation source. The position of tempered glass and scintillators is shown as a dotted line in front of the PMTs.}
    \label{fig:exp2_cartoon}
\end{figure}

The experimental setup is shown in Figure~\ref{fig:exp2_cartoon}.
A~60~mm in diameter stainless-steel guide was mounted to an UCN port, with a T-section to divert UCN to two separate detectors. Two identical 51~mm Hamamatsu R7724 PMTs were mounted at the end of each side, and elbow pieces were added in order reduce cross-talk between them. A tempered glass window was placed in front of the PMTs to protect them from low pressures in the UCN guide. A measurement for \YAP\ crystal scintillator as well as ZnS screen were performed. For ZnS, scintillator screen with 120~nm \B-coating on one side was optically coupled to the tempered glass using optical adhesive. \YAP\ measurements were done using the 120~nm \B-coated crystal from the previous section; it was placed directly on the tempered glass without any coupling. A gate valve (GV) at the UCN port could be closed to enable background measurements. Each measurement had two parts: first, a signal measurement with a GV open and UCN flow was enabled to the PMTs, and a background measurement, when the GV was closed to prevent UCN from getting to the detectors.
\subsection{Efficiency of \B\ capture}
The efficiency of the UCN \B-capture was established by a coincidence measurement between the \YAP\ signal from the PMT and a 478~keV gamma photon measured by a high-purity Germanium detector (HPGeD). For this measurement, only one of the PMTs in Figure~\ref{fig:exp2_cartoon} was used; the dashed PMT was removed and replaced with a stainless-steel end-cap. For both, the PMT and the HPGeD, the signal was amplified with 671 ORTEC spectroscopy amplifier and a pulse-height analysis was performed by FAST ComTec 7072 ADC with 500~ns time resolution. 

\begin{figure}
    \centering
    \includegraphics[width=\linewidth]{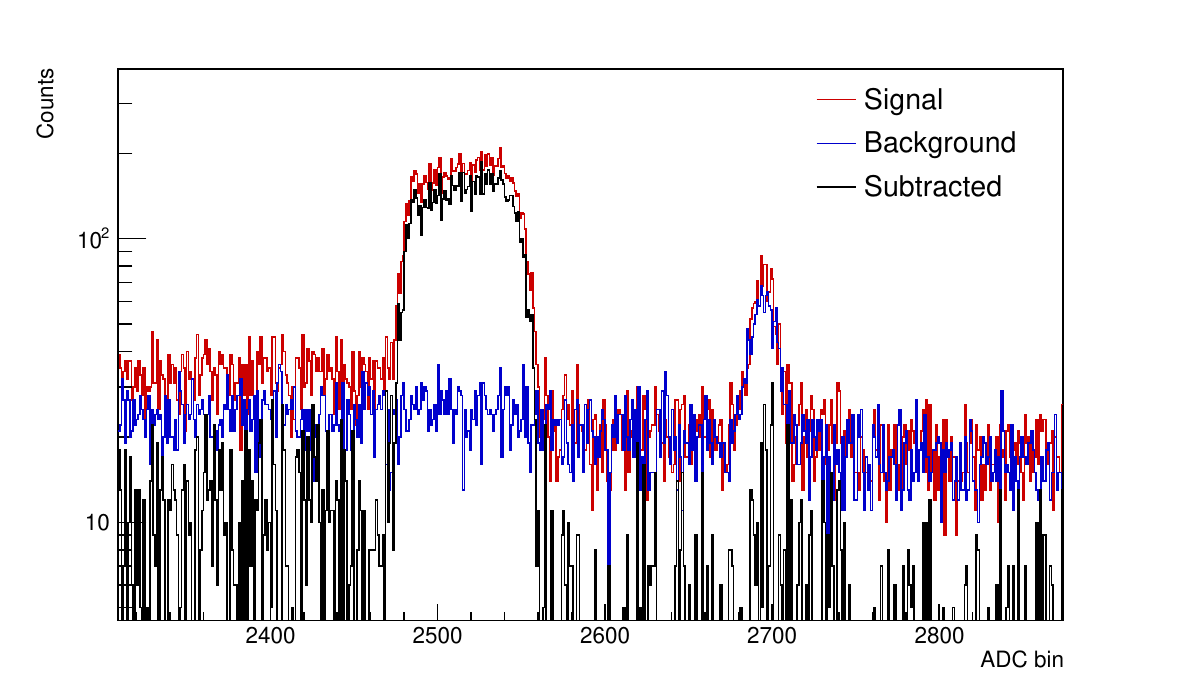}
    \caption{A gamma energy spectrum from the high-purity Germanium detector: signal (red) and background (blue) measurent with UCN GV open and closed, respectively. Their difference, i.e. a background-subtracted spectrum, is shown in black. Events above the background level near bin 2500 represent the 478~keV gammas from \B\ capture and events near bin 2700 correspond to 511~keV gammas.}
    \label{fig:YAP_crystal_Ge}
\end{figure}

\begin{figure}
    \centering
    \includegraphics[width=\linewidth]{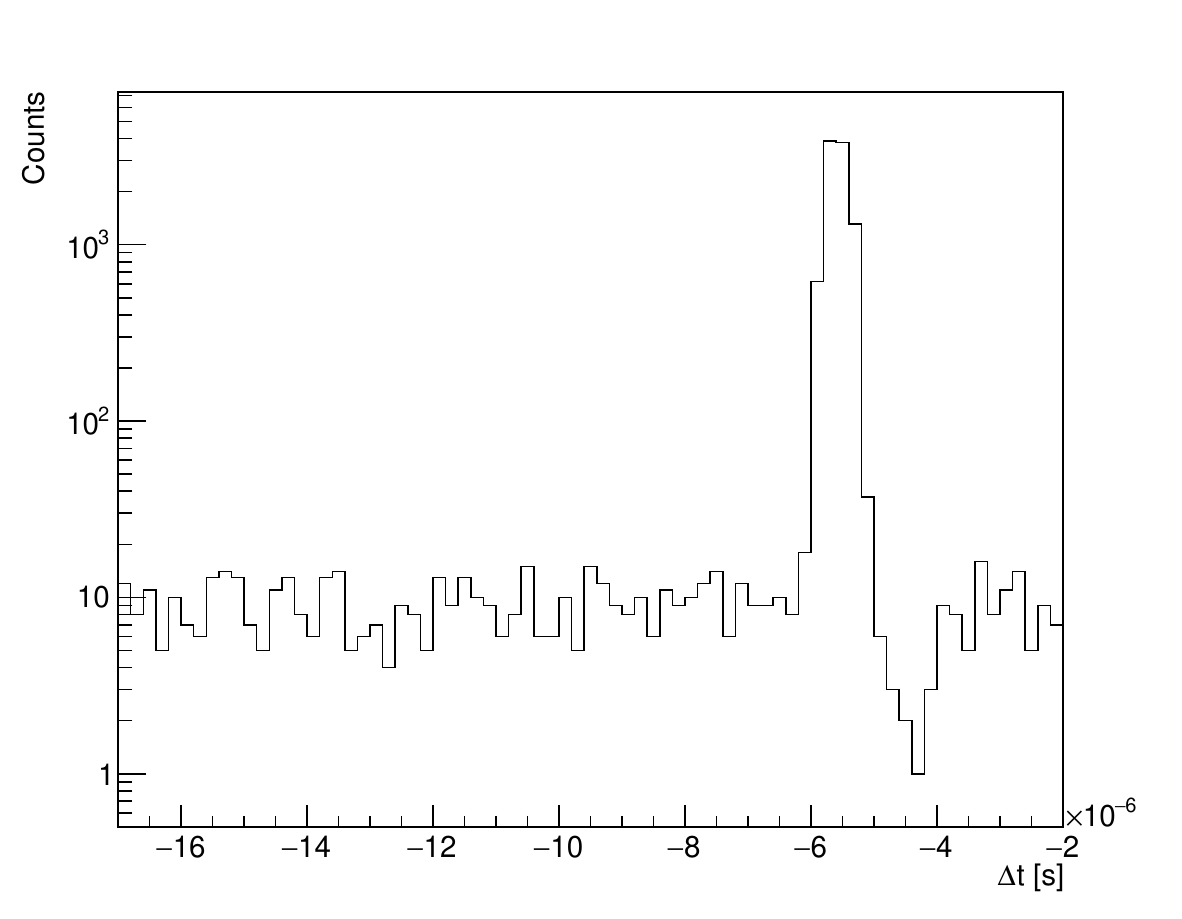}
    \caption{A coincidence between 478~keV gamma in HPGeD (at $t=0$) and a signal from \YAP.}
    \label{fig:coincidence_dt}
\end{figure}

The neutron detection efficiency, $\epsilon_\mathrm{N}$, from Equation~\ref{eq:lifetime} is estimated as:
\begin{equation}
    \epsilon_\mathrm{N} = \frac{N_\mathrm{PMT}}{N_\mathrm{Ge}},
    \label{eq:efficiency}
\end{equation}
where $N_\mathrm{PMT}$ and $N_\mathrm{Ge}$ are counts of UCNs detected by the PMT in coincidence with 478~keV gammas detected by the HPGeD and the total number of 478~keV gammas in HPGeD, respectively. Thirty minutes long runs were taken for \YAP\ crystal and ZnS screen. Figure~\ref{fig:YAP_crystal_Ge} shows HPGeD energy spectrum with 478~keV gamma peak; the background subtracted counts give $N_\mathrm{Ge}$ for Equation~\ref{eq:efficiency}. $N_\mathrm{PMT}$ was extracted from PMT counts in coincidence with the 478~keV gamma photons. Because HPGeD has a slower response than the PMT, the actual PMT signal arrives first into the ADC. The coincidence window spectrum from the signal run is shown in Figure~\ref{fig:coincidence_dt}, and events forming the peak at $\approx 5~\mu$s give $N_\mathrm{PMT}$. A constant background, fitted in the vicinity of the peak was subtracted from the number of counts. The same coincidence window was investigated for the background run with a result of no events; this means, that the background gamma photons (the blue spectrum in Figure~\ref{fig:YAP_crystal_Ge}) did not contribute to the $N_\mathrm{PMT}$. The choice of HPGeD window for coincidence triggers and the PMT coincidence time window were taken as a systematic uncertainty. The central values of the HPGeD coincidence window were bins 2470 to 2562, and the systematic uncertainty was evaluated by changing the window boundaries in both direction by 10 bins. The PMT time-coincidence window was $-(6.3; 4.3)~\mu$s for the central value and $-(7.1; 3.5)~\mu$s for the systematic uncertainty. The time window uncertainty has no appreciable effect. The \YAP\ efficiency, $\epsilon_N$, was measured to be $(86.82 \pm 2.61)\%$, and that for ZnS was $(91.59 \pm 3.36)\%$. The detailed uncertainties contribution is listed in Table~\ref{tab:efficiency_uncert}.
\begin{table}[h]
\begin{tabular}{cccc}
Scintillator & Statistical & HPGeD & Total \\ \hline
\YAP\ &  1.33 & 2.25 & 2.61 \\ 
ZnS & 1.64 & 2.93 & 3.36\\
\end{tabular}
\caption{Detail of uncertainties for neutron capture efficiency. From left statistical uncertainty and systematic uncertainty due to variation of 478~keV window in HPGeD.}
\label{tab:efficiency_uncert}
\end{table}
\subsection{\YAP\ as an absolute UCN counter}
Counting capabilities of \YAP\ scintillator were compared to the golden standard ZnS screen scintillator. By having the two scintillators measured at the same time, normalization is not required and their counts can be compared directly. Raw PMT signal was recorded by 10~ns resolution CAEN DT5724 digitizer; for each event, 20~$\mu$s long waveforms were recorded. Following methods used in the previous section, a constant offset, equal to an average signal in 500~ns before the signal was subtracted from the waveforms on event by events basis.

\begin{figure}
    \centering
    \includegraphics[width=\linewidth]{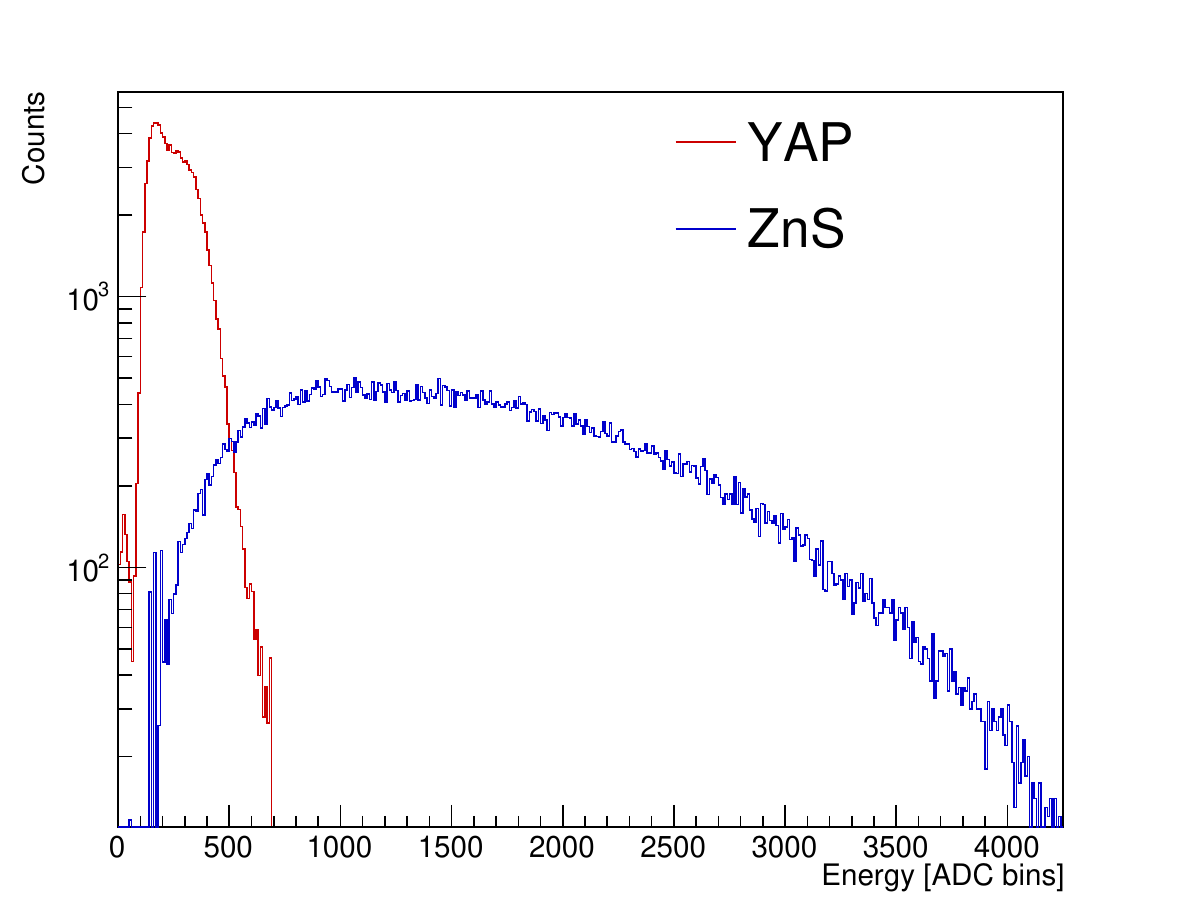}
    \caption{A background-subtracted energy spectrum of \YAP\ (red) and ZnS (blue) scintillator from the UCN \B\ capture.}
    \label{fig:YAP_ZnS}
\end{figure}

\begin{figure}
    \centering
    \includegraphics[width=\linewidth]{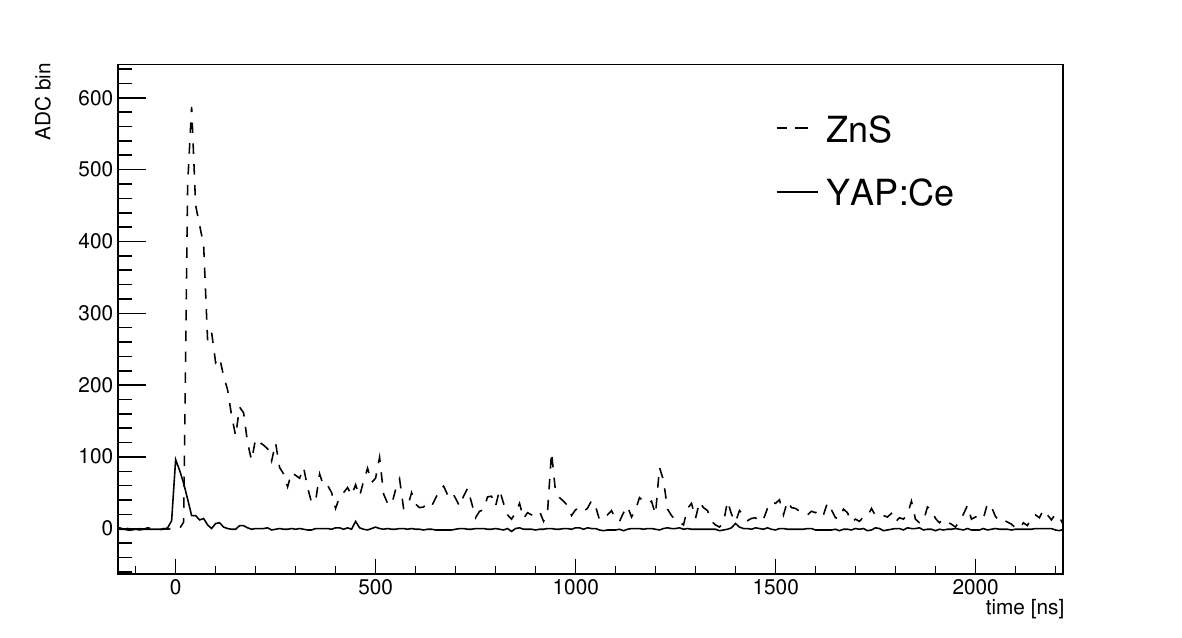}
    \caption{An exemplary waveform for one event from \YAP\ (solid) and one event from ZnS (dashed). For a better comparison, the horizontal axis was shifted so that the origin is approximately at waveforms rise time and the baseline offset was removed from both signals.}
    \label{fig:YAP_ZnS_waveform}
\end{figure}

10-minute long measurements were taken for both \YAP\ and ZnS scintillator.
Due to differences in counting rates, related PMT afterpulses, and incoming neutrons, background was scaled to fit the signal spectrum. Three Gaussian functions, representing capture of \Li, $\alpha$, and \Li+$\alpha$, with a background were fitted to the signal, resulting in a background scaling constant. The background-subtracted \YAP\ and ZnS energy spectrum is shown in Figure~\ref{fig:YAP_ZnS}. UCN counts are obtained by integrating the whole spectrum. An example of a waveform from both scintillators is shown in Figure~\ref{fig:YAP_ZnS_waveform}. Around 450~ns and 1200~ns in the \YAP\ spectrum, two single photoelectron (PE) signals are visible; using their waveform integral, \YAP\ waveform contains about 40~PEs and ZnS about 350~PEs. PMTs in the UCNProBe experiment will be used without the tempered glass, and their quantum efficiency peaks at \YAP's emission spectrum ($\approx380$~nm), which will further improve the light collection. 
The ratio of UCN counts, $R$, was evaluated as a ratio of corrected counts $N^\mathrm{corr}$:
\begin{equation}
  R = \frac{N^\mathrm{corr}_{\mathrm{YAP}}}{N^\mathrm{corr}_{\mathrm{ZnS}}}; \,\,\,\,\,  N^\mathrm{corr} = \frac{1}{A}\frac{N^{\mathrm{raw}}}{\epsilon_N},
    \label{eq:N_corrected}
\end{equation}
where $\epsilon_N$ is defined in Equation~(\ref{eq:efficiency}), and $N^{\mathrm{raw}}$ and $A$ are total uncorrected counts and the area of each scintillator, respectively. Systematic uncertainties were estimated by varying the integration time of the waveforms and changing the background scaling constant from fitted value to 1, i.e. no scaling. 
Because the equipment used in this and the previous section differ, two scenarios were considered. First, when $\epsilon_N$ was the same for both scintillators, and second, where values from the previous section were used. Using this conservative interval including $1\sigma$ uncertainties yields $(R-1) \in (8;28)\%$ , i.e. \YAP\ has outperformed ZnS by 8\% to 28\%. This is most likely caused by uneven \B-coating on ZnS' rough surface, compared to smooth surface of the \YAP\ crystal. Neutrons are not captured by a thinner or absent layer of \B\ and are upscattered in ZnS screen and lost. This makes \YAP\ crystal a better absolute UCN counter.

Future studies of \YAP\ in the actual UCNProBe chamber are expected to reduce the shortcomings of this measurement by using a thinner \B\ coating, PMTs with better quantum efficiency at \YAP's emission spectrum and overall better light collection.
\section{Results}
This work presents light properties from $\alpha$-particle radiation and UCN counting abilities of \YAP\ scintillator.

The measured Birks' coefficient for the \YAP\ scintillator is $(5.56 ^{+0.05}_{-0.30})\times 10^{-4}$~cm/MeV. About 50\% of the output light is lost if the light is to transmit through layers of deuterated polystyrene scintillator. The light losses due to 120~nm \B-coating were established to be about 60\%. A further study of a thinner \B\ coating is desired, as it can decrease the light losses. Our Birks' coefficient was successfully applied to $^{226}$Ra decay $\alpha$ particles presented in Ref.~\onlinecite{Baryshevsky1991}.

The \YAP\ scintillator produced less light output compared to ZnS screen. The efficiency of the \YAP\ scintillator as measured by 478~keV gamma coincidence was $(86.82 \pm 2.61)\%$. It is important to note that the inefficiency of the YAP scintillator is mostly due to the low light yield from $^7$Li ion. In this case, we expect the associated $\alpha$ to deposit its energy in the dPS for the final UCNProBe experiment, thus increasing the overall light output. Measurement with ultracold neutrons showed that \YAP\ crystal counted 8\% to 28\% more UCNs compared to ZnS screen. This may be due to uneven coating of \B\ on the rough surface.

In the final UCNProBe experiment, a much better light collection and therefore an even better efficiency is expected. This study has shown a novel concept of a UCN detector that can be calibrated absolutely. This scintillating crystal based Boron film detector has the advantage of short scintillating decay time and high Fermi potential. When accompanied with 478 keV gamma tagging, it allows $\alpha$/$^7$Li counting to be directly tied to UCN wall collision rate, without the complete understanding of UCN upscatttering, thus allowing absolute UCN counting at 0.1\% precision, assuming all other sources of UCN losses in \B\ coating and storage cell are at (typical) levels below $\approx10^{-3}$.
\bibliography{yap}
\end{document}